\documentclass{ws-procs9x6}

\begin{document}

\def\AJ{{\it Astrophys. J.} }
\def\AJL{{\it Ap. J. Lett.} }
\def\AJS{{\it Ap. J. Supp.} }
\def\AM{{\it Ann. Math.} }
\def\AP{{\it Ann. Phys.} }
\def\APJ{{\it Ap. J.} }
\def\ATP{{\it Astropart. Phys.} }
\def\APP{{\it Acta Phys. Pol.} }
\def\ASAS{{\it Astron. and Astrophys.} }
\def\BAMS{{\it Bull. Am. Math. Soc.} }
\def\CMJ{{\it Czech. Math. J.} }
\def\CMP{{\it Commun. Math. Phys.} }
\def\CQG{{\it Class. Quantum Gravity} }
\def\FP{{\it Fortschr. Physik} }
\def\GRG{{\it Gen. Relat. and Gravitation}}
\def\HPA{{\it Helv. Phys. Acta} }
\def\IJMP{{\it Int. J. Mod. Phys.} }
\def\JMM{{\it J. Math. Mech.} }
\def\JP{{\it J. Phys.} }
\def\JCP{{\it J. Chem. Phys.} }
\def\LNC{{\it Lett. Nuovo Cimento} }
\def\SNC{{\it Suppl. Nuovo Cimento} }
\def\MNRAS{{\it Mont. Not. R. Astr. Soc.} }
\def\MNRASL{{\it Mont. Not. R. Astr. Soc. Lett.} }
\def\MPL{{\it Mod. Phys. Lett.} }
\def\NAT{{\it Nature} }
\def\NC{{\it Nuovo Cimento} }
\def\NP{{\it Nucl. Phys.} }
\def\PM{{\it Philos. Mag. } }
\def\PL{{\it Phys. Lett.} }
\def\PR{{\it Phys. Rev.} }
\def\PRL{{\it Phys. Rev. Lett.} }
\def\PRTS{{\it Physics Reports} }
\def\PS{{\it Physica Scripta} }
\def\PTP{{\it Progr. Theor. Phys.}}
\def\RPP{{\it Rep. Progr. Phys.}}
\def\RMPA{{\it Rev. Math. Pure Appl.} }
\def\RMP{{\it Rev. Mod. Phys.} }
\def\RNC{{\it Rivista del Nuovo Cimento} }
\def\UFN{{\it Usp. Fiz. Nauk} }
\def\SJPN{{\it Soviet J. Part. Nucl.} }
\def\SPU{{\it Soviet. Phys. Usp.} }
\def\TMF{{\it Teor. Mat. Fiz.} }
\def\TMP{{\it Theor. Math. Phys.} }
\def\YF{{\it Yadernaya Fizika} }
\def\ZETF{{\it Zh. Eksp. Teor. Fiz.} }
\def\ZP{{\it Z. Phys.} }
\def\ZMP{{\it Z. Math. Phys.} }
\def\NCA{\em Nuovo Cimento}
\def\NIM{\em Nucl. Instrum. Methods}
\def\NIMA{{\em Nucl. Instrum. Methods} A}
\def\ZPC{{\em Z. Phys.} C}
\def\st{\scriptstyle}
\def\sst{\scriptscriptstyle}
\def\mco{\multicolumn}
\def\epp{\epsilon^{\prime}}
\def\vep{\varepsilon}
\def\ra{\rightarrow}
\def\ppg{\pi^+\pi^-\gamma}
\def\vp{{\bf p}}
\def\ko{K^0}
\def\kb{\bar{K^0}}
\def\al{\alpha}
\def\ab{\bar{\alpha}}
\def\beq{\begin{equation}}
\def\eeq{\end{equation}}
\def\beqa{\begin{eqnarray}}
\def\eeqa{\end{eqnarray}}
\def\CPbar{\hbox{{\rm CP}\hskip-1.80em{/}}}

\def\al{\alpha}
\def\be{\beta}
\def\ga{\gamma}
\def\de{\delta}
\def\ep{\epsilon}
\def\ve{\varepsilon}
\def\ze{\zeta}
\def\et{\eta}
\def\th{\theta}
\def\vt{\vartheta}
\def\io{\iota}
\def\ka{\kappa}
\def\la{\lambda}
\def\vpi{\varpi}
\def\rh{\rho}
\def\vr{\varrho}
\def\si{\sigma}
\def\vs{\varsigma}
\def\ta{\tau}
\def\up{\upsilon}
\def\ph{\phi}
\def\vp{\varphi}
\def\ch{\chi}
\def\ps{\psi}
\def\om{\omega}
\def\Ga{\Gamma}
\def\De{\Delta}
\def\Th{\Theta}
\def\La{\Lambda}
\def\Si{\Sigma}
\def\Up{\Upsilon}
\def\Ph{\Phi}
\def\Ps{\Psi}
\def\Om{\Omega}
\def\mn{{\mu\nu}}
\def\prt{\partial}
\def\ap{\al^\prime}
\def\apt{\al^{\prime 2}}
\def\apth{\al^{\prime 3}}
\def\pt#1{\phantom{#1}}
\def\vev#1{\langle {#1}\rangle}
\def\bra#1{\langle{#1}|}
\def\ket#1{|{#1}\rangle}
\def\bracket#1#2{\langle{#1}|{#2}\rangle}
\def\expect#1{\langle{#1}\rangle}
\def\sbra#1#2{\,{}_{{}_{#1}}\langle{#2}|}
\def\sket#1#2{|{#1}\rangle_{{}_{#2}}\,}
\def\sbracket#1#2#3#4{\,{}_{{}_{#1}}
 \langle{#2}|{#3}\rangle_{{}_{#4}}\,}
\def\sexpect#1#2#3{\,{}_{{}_{#1}}\langle{#2}\rangle_{{}_{#3}}\,}
\def\half{{\textstyle{1\over 2}}}
\def\frac#1#2{{\textstyle{{#1}\over {#2}}}}
\def\ni{\noindent}
\def\lsim{\mathrel{\rlap{\lower4pt\hbox{\hskip1pt$\sim$}}
    \raise1pt\hbox{$<$}}}
\def\gsim{\mathrel{\rlap{\lower4pt\hbox{\hskip1pt$\sim$}}
    \raise1pt\hbox{$>$}}}
\def\sqr#1#2{{\vcenter{\vbox{\hrule height.#2pt
	 \hbox{\vrule width.#2pt height#1pt \kern#1pt
	 \vrule width.#2pt}
	 \hrule height.#2pt}}}}
\def\square{\mathchoice\sqr66\sqr66\sqr{2.1}3\sqr{1.5}3}
\def\Re{\hbox{Re}\,}
\def\Im{\hbox{Im}\,}
\def\Arg{\hbox{Arg}\,}
\def\z{{\bf\hat z}}
\def\cD{{\cal D}}
\def\cL{{\cal L}}
\def\cH{{\cal H}}
\def\va{\vev{A}}
\def\vb{\vev{B}}
\def\vc{\vev{C}}
\def\oc{\widehat{\cos\ph}}
\def\os{\widehat{\sin\ph}}

\title{Dark Energy - Dark Matter Unification: \\
Generalized Chaplygin Gas Model\footnote{\uppercase{T}alk presented at 
the \uppercase{V} \uppercase{N}ew \uppercase{W}orlds in \uppercase{A}stroparticle \uppercase{P}hysics \uppercase{C}onference, 
\uppercase{F}aro, \uppercase{P}ortugal, 8-10 \uppercase{J}anuary 2005.}}

\author{ORFEU BERTOLAMI}

\address{Instituto Superior T\'ecnico, Departamento de F\'\i sica\\
Av. Rovisco Pais, 1049-001\\ 
Lisboa, Portugal\\ 
E-mail: orfeu@cosmos.ist.utl.pt}

\maketitle

\abstracts{
We review the main features of the generalized Chaplygin gas (GCG) proposal for
unification of dark energy and dark matter and discuss how it admits an
unique decomposition into dark energy and dark matter components once
phantom-like dark energy is excluded. In the context of this approach 
we consider structure formation and show that 
unphysical oscillations or blow-up in the matter power spectrum 
are not present. Moreover, we demonstrate
that the dominance of dark energy occurs about the time when energy  
density fluctuations start evolving away from the linear regime. }

\section{Introduction}

The GCG model\cite{Kamenshchik,Bento1} is an interesting alternative 
to more conventional approaches for explaining the observed accelerated
expansion of the Universe such as a
cosmological constant\cite{Bento2} or quintessence\cite{Ratra}. It is 
worth remarking that quintessence is related to 
the idea that the cosmological term could evolve\cite{Bronstein} and with attempts to tackle the 
cosmological constant problem. 

In the GCG approach one considers an exotic equation of state 
to describe the background fluid:

\beq
p_{ch} = - {A \over \rho_{ch}^{\alpha}} ~~,
\label{eqstate}
\eeq
where $A$ and $\alpha$ are positive constants. The case $\alpha=1$
corresponds to the Chaplygin gas. In most phenomenological studies 
the range $0 < \alpha \le 1$ is considered. Within the
framework of Friedmann-Robertson-Walker cosmology, this equation of state
leads, after being inserted into the relativistic energy conservation equation, to
an evolution of the energy density as\cite{Bento1}

\beq
\rho_{ch}=  \left[A + {B \over a^{3 (1 + \alpha)}}\right]^{1 \over 1 +
\alpha}~~,
\label{rhoch}
\eeq
where $a$ is the scale-factor of the Universe and $B$ a positive
integration constant. From this result, one can understand a striking property of the GCG: 
at early times the energy density behaves as matter while at late times it behaves
like a cosmological constant. This dual role is what essentially allows for the interpretation of 
the GCG model as an entangled mixture of dark matter and dark energy.

The GCG model has been successfully confronted with different classes of 
phenomenological tests: high precision Cosmic Microwave Background
Radiation data\cite{Bento3}, supernova data\cite{Supern}, and
gravitational lensing\cite{Silva}. More recently, 
it has been shown using the latest supernova data\cite{Tonry}, 
that the GCG model is degenerate with a dark energy model with 
a phantom-like equation of state\cite{Bertolami1,Bento4}. Furthermore, it can be shown that this 
does not require invoking the unphysical condition of violating the 
dominant energy condition and does not lead to the big rip singularity in future\cite{Bertolami1}. 
It is a feature of GCG model, that it can mimic a phantom-like equation of state, 
but without any kind of pathologies 
as asymptotically the GCG approaches to a well-behaved de-Sitter
universe. Structure formation has been studied in Refs. [2, 12]. 
In Ref. [13], the results of the various 
phenomenological tests on the GCG model are summarized.

Despite these pleasing performance concerns about such an
unified model were raised in the context of structure formation. Indeed, it 
has been pointed out that one should expect unphysical oscillations or even an
exponential blow-up in the matter power spectrum at present\cite{Sandvik}.  
This difficult arises from the behaviour of the sound
velocity through the GCG. Although, at  early times, the GCG
behaves like  dark matter and its sound velocity is vanishingly small as one
approaches the present, the GCG starts behaving like dark energy
with a substantial negative pressure yielding a large sound velocity
which, in turn,  produces oscillations or blow-up in the power
spectrum. In any unified approach this is inevitable unless 
the dark matter and the dark energy components of the fluid can be properly identified. 
These components are, of course, interacting as both are
entangled within a single fluid. However, it can be shown that the GCG is a unique mixture of
interacting dark matter and a cosmological constant-like dark energy,
once one excludes the possibility of phantom-type dark energy\cite{Bento5}. 
It can be shown that due to
the interaction between the components, there is a flow of energy from
dark matter to dark energy. This energy
transfer is vanishingly small until recent past, resulting in a
negligible contribution at the time of
gravitational collapse ($z_c \simeq 10$). This feature makes the model
indistinguishable from a CDM dominated Universe till recent past. Subsequently, 
just before present ($z \simeq 2$), the interaction starts to
grow yielding a large energy transfer from dark matter to dark
energy, which leads to the dominance of the latter at present. Moreover, it is shown that
the epoch of dark energy dominance occurs when dark
matter perturbations start deviating from its linear behaviour
and that the Newtonian equations for small scale
perturbations for dark matter do not involve any $k$-dependent
term. Thus, neither oscillations nor blow-up in the power spectrum do develop.


\section{Decomposition of the GCG fluid}


In Ref. [2], it is shown that the GCG can be described through a 
complex scalar field whose Lagrangian density can be 
written as a \textit{generalized} Born-Infeld theory:

\beq
{\it L}_{GBI} = - A^{1 \over 1 + \alpha} 
\left[1 - (g^{\mu \nu} \theta_{, \mu} 
\theta_{, \nu})^{1 + \alpha \over 2\alpha}\right]^{\alpha \over 1 + \alpha}~~,
\label{GenBornInfeld}
\eeq
which reduces into the Born-Infeld Lagrangian density for
$\alpha = 1$. The field $\theta$ corresponds to the phase of the complex
scalar field\cite{Bento1}.

Let us now consider the decomposition of the GCG into components. Introducing 
the redshift dependence and using Eqs. (\ref{eqstate}) and (\ref{rhoch}), the pressure is given by

\beq
p_{ch} = -{A \over \left[A+B(1+z)^{3(1+\alpha)}\right]^{\frac{\alpha}
{1+\alpha}}}
\label{totalp}
\eeq
while the total energy density can be written as

\beq
\rho_{ch}=\left[A+B(1+z)^{3(1+\alpha)}\right]^{\frac{1}{1+\alpha}}~~, 
\label{totalrho}
\eeq
where the present value of the scale-factor, $a_0$, has been set to $1$.

We decompose the energy density into a pressure-less dark matter
component, $\rho _{dm}$, and a dark energy component, $\rho _{X}$, with
an equation of state $p_X = w_X \rho_{X}$; hence the equation of state
parameter of the GCG can be written as

\beq
w={p_{ch} \over \rho_{ch}}={p_X \over \rho_{dm}+\rho_X}={w_X\rho_X
  \over 
\rho_{dm}+\rho_X}~~.
\label{espar}
\eeq

Therefore, from Eqs. (\ref{totalp}), (\ref{totalrho}) and (\ref{espar}), one
obtains for $\rho_X$

\beq
\rho_X=-{ \rho_{dm} \over
 1+w_X\left[1+\frac{B}{A}(1+z)^{3(1+\alpha)}\right]}~~.
\label{rhox}
\eeq

From this equation one can see that requiring that $\rho _{X} \ge 0$ leads to the
constraint $w_{X} \le 0$ for early times ($z\gg 1$) and $w_{X} \le -1$
for the future ($z=-1$). Thus, one can conclude that $w_{X} \le -1$ for the
entire history of the Universe.  The case $w_{X}<-1$ corresponds to
the so-called phantom-like dark energy, which violates the
dominant-energy condition and leads to an ill defined sound velocity
(see however Ref. [10]). Excluding this possibility, then
the energy density can be uniquely split as

\begin{equation}
\rho=\rho_{dm}+\rho_\Lambda  \label{split}
\end{equation}
where

\beq \rho_{dm}={B (1+z)^{3(1 + \alpha)} \over \left[A + B
(1+z)^{3(1+\alpha)}\right]^{\frac{\alpha}{1+\alpha}}}~~,
\label{rhom}
\eeq
and

\beq \rho_\Lambda=-p_\Lambda={A \over
\left[A+B(1+z)^{3(1+\alpha)}\right]^{\frac{\alpha}{1+\alpha}}}~~,
\label{rholam}
\eeq
from which one finds the scaling behaviour of the energy densities

\beq
{\rho_{dm} \over \rho_\Lambda}={B \over A} (1+z)^{3(1+\alpha)}~~.
\label{scale}
\eeq

\begin{figure}[ht]
\begin{center}
\epsfxsize 2.8in
\epsfbox{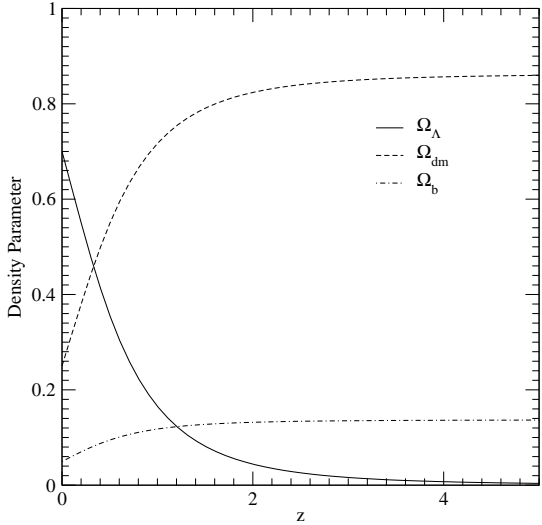}
\caption{\label{fig:fig1} Density parameters 
$\Omega_{dm}$ and $\Omega_\Lambda$ and
$\Omega_{b}$ as a function of redshift. It is assumed that $\Omega_{b0}=0.05$, 
$\Omega_{dm0}=0.25$, $\Omega_{\Lambda0}=0.7$ and $\alpha =0.2$.}
\end{center}
\end{figure}

In what follows we express parameters A and B in terms of
cosmological observables. From Eqs. (\ref{rhom}) and (\ref {rholam}),
it implies that

\beq
\rho_{ch0}=\rho_{dm0}+\rho_{\Lambda 0}= (A+B)^{\frac{1}{1+\alpha}}~~,
\label{rel}
\eeq 
where $\rho _{ch0}$, $\rho _{dm0}$ and $\rho_{\Lambda 0}$ 
are the present values of $\rho_{ch}$, $\rho_{m}$ and
$\rho _{\Lambda}$, respectively. Constants $A$ and $B$ can then be
written as a function of $\rho _{ch0}$

\beq
A=\rho_{\Lambda 0}~\rho_{ch0}^\alpha~; \quad
B=\rho_{dm0}~\rho_{ch0}^\alpha~~.
\label{AB}
\eeq

It is also interesting to express $A$ and $B$ in terms of $\Omega _{dm0}$,
$\Omega_{\Lambda 0}$, the present values of the fractional energy
densities $\Omega _{dm(\Lambda )}={\rho _{m(\Lambda )}/\rho _{c}}$
where $\rho _{c}$ is the critical energy density, $\rho _{c}= 3H^2/8
\pi G$. From the Friedmann equation

\beq
3H^2  = 8 \pi G \left[A + B 
(1+z)^{3(1+\alpha)}\right]^{\frac{1}{1+\alpha}}
+ 8\pi G\rho_{b0}(1+z)^3
\label{h2}
\eeq
where $\rho_{b0}$ is the baryon energy density at present, one obtains

\beq
A \simeq \Omega_{\Lambda 0}~ \rho_{c 0}^{(1+\alpha)}, ~
B \simeq \Omega_{dm 0}~ \rho_{c 0}^{(1+\alpha)}~~.
\label{params}
\eeq

Therefore, with the present value of the Hubble parameter, $H_0$: 

\beq
H^2= H_0^2 \left[\left[\Omega_{\Lambda 0} + \Omega_{dm 0} (1+z)^{3(1+\alpha)}
\right]^{\frac{1}{1+\alpha}} + \Omega_{b0}(1+z)^{3}\right]~~
\label{hubble2}
\eeq
one can write the fractional energy
densities $\Omega _{dm},~\Omega _{\Lambda }$  and $\Omega_{b}$ as

\beq
\Omega_{dm} =  {\Omega_{dm0}(1+z)^{3(1+\alpha)} \over
\left[\Omega_{\Lambda
    0}+\Omega_{dm0}(1+z)^{3(1+\alpha)}\right]^{\alpha/
(1+\alpha)}X}
\eeq
\beq
\Omega_\Lambda  = {\Omega_{\Lambda 0} \over
\left[\Omega_{\Lambda 0}+\Omega_{dm0}(1+z)^{3(1+\alpha)}\right]^{\alpha/
(1+\alpha)}X}
\eeq
\beq
\Omega_b = {\Omega_{b0}(1+z)^{3} \over X}
\eeq
where
\beq
 X=\left[\Omega_{\Lambda
 0}+\Omega_{m0}(1+z)^{3(1+\alpha)}\right]^{1/(1+\alpha)}+\Omega_{b0}(1+z)^3~.
\eeq

Finally, as $\Omega_{dm0}$ and $\Omega_{\Lambda 0}$ are order
one quantities, one can easily see that at the time of nucleosynthesis, 
$\Omega_\Lambda$ is negligibly small, and hence the model is not in conflict 
with known processes at nucleosynthesis.

It is important to realize that there is an explicit interaction between dark matter and
dark energy. This can be understood from the energy conservation equation,
which in terms of the components can be written as

\beq
\dot \rho_{dm} + 3 H \rho_{dm} = - \dot \rho_\Lambda ~~.
\label{evorho}
\eeq
Thus, the evolution of dark energy and dark matter are coupled so
that energy is exchanged between these components 
(see Refs. [16,17] for earlier work on the interaction
between dark matter and dark energy). One can see from Figure 1, that
until $z \simeq 2$, there is essentially no exchange of energy and the
$\Lambda$ term is vanishingly small.  However, around $z \simeq 2$, the
interaction starts to increase, resulting in a substantial growth of the
dark energy term at the expense of the dark matter energy. 
Thus, by around $z \simeq 0.2$, dark energy starts dominating the
energy content of Universe. Of course, these redshift values are $\alpha$ dependent 
and, in Figure 1,
$\alpha = 0.2$ has been chosen. Nevertheless, the main conclusion is that in this
unified model, the interaction between dark matter and dark
energy is vanishing small for almost the entire history of the
Universe making it indistinguishable from the CDM model.  As can be clearly seen, the energy
transfer has started in the recent past resulting in a significant
energy transfer from dark matter to the $\Lambda$-like dark energy. In
the next section we show that this energy transfer epoch is the 
one when dark matter perturbations start departing from
its linear behaviour. But before that 
notice also that Eq. (\ref{evorho}) expresses the energy conservation for the background fluid, which   
is reminiscent of earlier work on varying $\Lambda$
cosmology\cite{Bronstein,Freeze,Waga} where the cosmological term
decays into matter particles. In here, we have the opposite, as
$\alpha$ is always positive, hence the energy transfer is from dark
matter to dark energy. This is responsible for the late time dominance of the
latter and ultimately to the observed accelerated expansion of the
Universe.


\section{ Structure Formation}

Aiming to study structure formation, it is interesting to write the
0-0 component of  Einstein's equation as

\beq
3H^{2} = 8\pi G (\rho_{dm} + \rho_{b})+ \Lambda ~~,
\label{G00}
\eeq
where $\Lambda$ is given by

\beq
\Lambda = 8\pi G \rho_{\Lambda}~~.
\eeq

\begin{figure}[t]
\begin{center}
\epsfxsize 2.8in 
\epsfbox{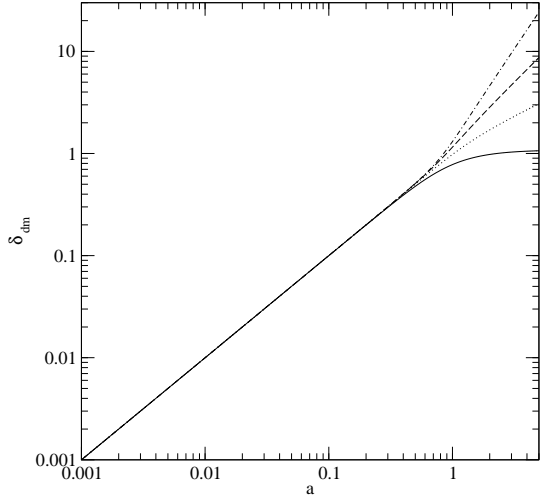}
\caption{Density profile 
$\delta_{dm}$ as function of scale factor.  The solid,
dotted, dashed and dash-dot lines correspond to $\alpha = 0, 0.2, 0.4,
0.6$, respectively. It is assumed that $\Omega_{b0}=0.05$, $\Omega_{dm0}=0.25$ and $\Omega_{\Lambda0}=0.7$.}
\end{center}
\end{figure}

We address now the issue of energy density perturbations. 
We first write the Newtonian equations for a pressure-less fluid
with background density $\rho_{dm}$ and density contrast
$\delta_{dm}$, with a source term due to the energy transfer from dark
matter to dark energy. Assuming that
both, the density contrast $\delta_{dm}$ and the peculiar velocity $v$ are
small, that is $\delta_{dm} <<1$ and $v << u$, where $u$ is the
velocity of a fluid element, one can write the Euler,
the continuity and the Poisson's equations in the co-moving frame\cite{Waga}:

\beqa 
\label{eq1}
\ddot{a}x + {\partial{v}\over{\partial{t}}} + {\dot{a}\over{a}}v
= - {\nabla\Phi\over{a}} ~~,\\ 
\label{eq2}
\nabla\cdot v =
-a\left[{\partial{\delta_{dm}}\over{\partial{t}}} +
{\Psi\delta_{dm}\over{\rho_{dm}}}\right] ~~,\\
\label{eq3}
{1\over{a^2}}\nabla^{2}\Phi = 4\pi G \rho_{dm}(1+\delta_{dm}) -
\Lambda ~~, 
\eeqa 
where $\Phi$ is the gravitational potential, and
$\Psi$ is the source term in the continuity equation due to the energy
transfer between dark matter and the cosmological constant-type dark
energy. The co-moving coordinate $x$ is related to the proper
coordinate $r$ by $r = ax$. In here, 

\beq 
\Psi = -{1\over{8\pi G}} \dot{\Lambda}~~.  
\eeq

One expects a perturbation also in the $\Lambda$ term. 
However, it can be seen from the  Euler equation, for a fluid with an equation state of the form $p=w \rho$,

\beq
(w+1)\rho\left({\partial v \over \partial t} + v\cdot \nabla v\right)
 + w\nabla \rho + (w+1)\rho \nabla \Phi=0
\eeq
so that, for $w=-1$, it follows that $\nabla \rho=0$, from which implies that this
cosmological constant like component is always homogeneous. We should mention that 
the Euler Eqs. (26) and (28) can  have an extra term in the r.h.s. if the 
velocity of the created $\Lambda$-like particle has a different velocity from the decaying dark 
matter particle\cite{Waga}. In this case, the $\Lambda$-like dark energy can have 
spatial variations which can be neglected for the Newtonian treatment. 
However, in our case, we are considering only the 
situation where both the decaying and created particles have the same velocity.

From the divergence of Eq. (\ref{eq1}) and using Eqs. (\ref{eq2}) and
(\ref{eq3}), one obtains the small scale linear perturbation
equation for the dark matter in the Newtonian limit:

\beq
{\partial^{2}\delta_{dm}\over{\partial t^2}} +
\left[2{\dot{a}\over{a}} + {\Psi\over{\rho_{dm}}}\right]
{\partial\delta_{dm}\over{\partial t}} - \left[4\pi
G\rho_{dm} - 2{\dot{a}\over{a}}{\Psi\over{\rho_{dm}}} -
{\partial\over{\partial t}}
\left[{\Psi\over{\rho_{dm}}}\right]\right]\delta_{dm} = 0 ~~.  
\eeq

One sees that, if $\Psi = 0$, that is in the absence of energy transfer,
one recovers the standard equation for the dark matter perturbation in the
$\Lambda$CDM case. One can verify that this occurs for $\alpha
=0$. It can also be seen from the above equation that there
is no scale dependent term to drive oscillations or to cause any blow up in the
power spectrum.

\begin{figure}[t]
\begin{center}
\epsfxsize 2.8in 
\epsfbox{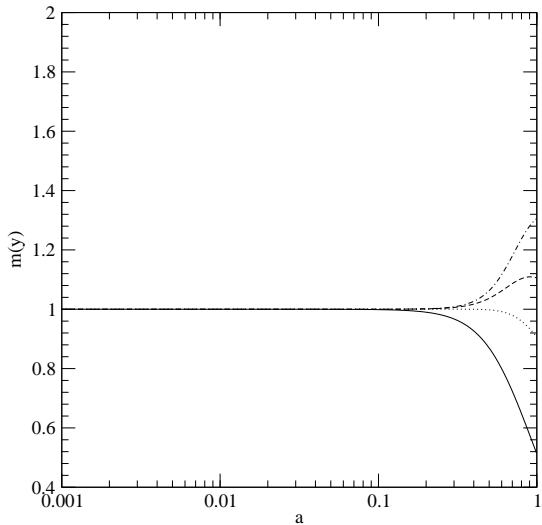}
\caption{The growth factor $m(y)$ as a function of scale factor a.
The solid, dotted, dashed and dash-dot lines correspond to $\alpha =
0, 0.2, 0.4, 0.6$, respectively. It is assumed that $\Omega_{b0}=0.05$, $\Omega_{dm0}=0.25$ 
and $\Omega_{\Lambda0}=0.7$.}
\end{center}
\end{figure}

We turn now to the evolution for the baryon perturbations in the
Newtonian limit when the scales are inside the horizon. Given that our purpose is to
consider the period after decoupling, the
baryons are no longer coupled to photons and one can effectively consider
baryons as a pressure-less fluid like the dark matter as there is no significant
pressure due to Thompson scattering. We assume that there is no interaction between 
dark energy and baryons, which means that the Equivalence Principle is violated 
as, on its turn, dark matter and dark matter are strongly coupled. 
Given that it is parameter $\alpha$ that controls this
interaction ($\alpha =0$ means there is no interaction), it is a
measure of the violation of the Equivalence Principle. One can also see 
from the behaviour of $\Psi$, that this violation also starts
rather late in the history of the Universe. In the Newtonian limit,
the evolution of the baryon perturbation after decoupling for scales
well inside the horizon is similar to the one for dark
matter however, as described earlier, the source term is absent as there is no energy transfer to
or from baryons. Thus, the equation for the evolution of baryon perturbations is given by

\beq {\partial^{2}\delta_{b}\over{\partial t^2}} +
2{\dot{a}\over{a}}{\partial\delta_{b}\over{\partial t}} - 4\pi
G\rho_{dm}\delta_{dm} = 0 ~~, 
\eeq 
where in the third term in the l.h.s., 
the contribution from baryons has been dropped as it is negligible
compared to the one of dark matter. 

It is convenient to define for each
component the linear growth function $D(y)$,

\beq 
\delta = D(y)\delta_{0} ~~, 
\eeq 
where $y=\log(a)$ and $\delta_{0}$ is the
initial density contrast (assuming a Gaussian distribution). It is also interesting to 
consider the so-called growth exponent $m(y) = D^{'}(y)/D(y)$, where the prime denotes derivative with 
respect to the scale factor.

Asymptotically, given that dark matter drives the evolution of the baryon
perturbations, then they grow with the same exponent $m(y)$. However,
their amplitudes may differ and their ratio corresponds to the
so-called  bias parameter, $b \equiv \delta_{b}/ \delta_{dm}$.

 
\begin{figure}[ht]
\begin{center}
\epsfxsize 2.8in 
\epsfbox{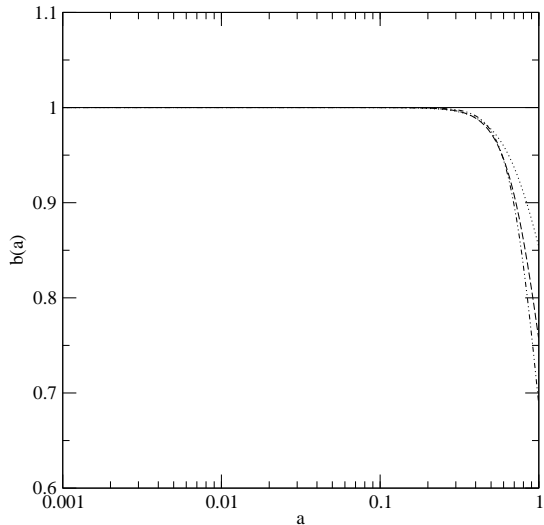}
\caption{Bias parameter $b$ as a function of the scale factor, $a$. The solid, 
dotted, dashed and dash-dot lines correspond to $\alpha = 0, 0.2, 0.4,
0.6$, respectively. It is assumed that $\Omega_{b0}=0.05$, $\Omega_{dm0}=0.25$ and 
$\Omega_{\Lambda0}=0.7$.}
\end{center}
\end{figure}

It is of course, phenomenologically interesting to 
study the behaviour of $\delta_{dm}$, $m(y)$ and
$b$ as function of the scale factor $a$.  While solving the
differential equations for the linear perturbation, the initial
conditions are chosen so that at  $a= 10^{-3}$, the standard linear
solution $D \simeq a$ is reached. In Figure 2, it is shown the
linear density perturbation for dark matter, $\delta_{dm}$, as a
function of $\alpha$.  One sees that, whereas for $\alpha = 0$ (the
$\Lambda$CDM case), the perturbation stops growing at late times, for
models with $\alpha > 0$ the
perturbation starts departing from the linear behaviour around 
$z \simeq 0.25$, the very epoch when the
$\Lambda$ term starts dominating (cf. Figure 1).  In view of this
behaviour, it is tempting to conjecture that, in our unified model,
the interaction between dark matter and $\Lambda$-like dark energy is
related with structure formation, so that for a sufficiently high density contrast
($\delta_{dm} >> 1$), a significant
energy transfer from dark matter to dark energy takes place. 
In any case, our proposal for GCG indicates that there is a  connection between
structure formation scenario and the dominance of dark energy, a link 
that ultimately results in the acceleration of the Universe
expansion. This feature hints a possible way to understand why 
$\Omega_{dm} \simeq \Omega_{\Lambda}$ just at recent past, the so-called 
Cosmic Coincidence problem.

 
\begin{figure}[t]
\begin{center}
\epsfxsize 2.8in 
\epsfbox{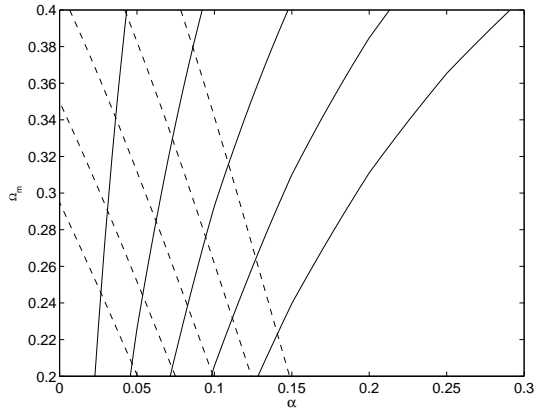}
\caption{Contours for parameters $b$ and $m$ in the $\Omega_m$-$\alpha$ plane.
Solid lines refer to $b$ whereas dashed lines refer to $m$. 
For $b$, contour values 
are 0.98, 0.96, ..., 0.9 from left to right. 
For $m$, contour values are 0.6, 0.65, ..., 0.8 from left to right.}
\end{center}
\end{figure}

The behaviour of $m(y)$ is also interesting. 
One can infer from Figure 3 that from $z \simeq 5$ to the present, 
the growth factor is quite sensitive to the value of
$\alpha$.  For $\alpha = 0.2$, $m(y)$ increases up to $40\%$ at present
in relation to the $\Lambda$CDM case. 

Notice that
$m(y)$ governs the growth of the velocity fluctuations in the linear
perturbation theory as the velocity divergence evolves as $-
H a m \delta_{dm}$; it follows then that large deviations of the growth factor
with changing $\alpha$ are detectable via precision measurements of 
large scale structure and associated measurements of the redshift-space 
power spectrum anisotropy.

In what concerns the bias parameter, its behaviour is shown in
Figure 4. From there one can see that it also changes sharply in the recent past as 
$\alpha$ increases. This bias extends to all scales consistent with the
Newtonian limit, hence being  distinguishable from the hydrodynamical
or nonlinear bias which takes place only for collapsed objects. Therefore,
from the observation of large scale clustering one can distinguish
the non-vanishing $\alpha$ case from the $\alpha = 0$ ($\Lambda$CDM) case.

The growth factor and the bias parameter at $z \sim 0.15$ have been recently
determined using the 2DF survey\cite{Haw,Verd}. It is found for the
redshift space distortion parameter, $\beta = 0.49 \pm 0.09$,
and for the linear bias, $b = 1.04 \pm 0.14$. Notice that,
as $\beta = m/b$, one can obtain $m =0.51 \pm 0.11$. 
In Figure 5, it is shown contours for $b$ and $m$ in the 
$\Omega_m$-$\alpha$ plane.
From the mentioned observational constraints on $b$ and $m$, one can 
constrain $\alpha$ to a small but non-zero value ($\alpha \sim 0.1$).
However, it is important to point out that
our study refers to the properties of the baryons whereas the
observations concern the fraction of baryons that collapsed to form
bright galaxies; the relation between the two is still poorly known. 
As far as parameter $\beta$ is
concerned, one should bear in mind that this constraint is obtained
in the context of the standard $\Lambda$CDM model in order to 
convert redshift to distance. Thus, a full analysis in the context of 
the GCG model is still to be performed.

Furthermore, as can be seen from Figure 2, there is no suppression of
$\delta_{dm}$ at late times for any positive value of $\alpha$, and
hence one should not expect the corresponding suppression in the power
spectrum normalization, $\sigma_{8}$, for the total matter
distribution. This was one major problem in the previous GCG model
approach which, as pointed out in Ref. [14],
cannot be solved even after the inclusion of baryons. In the approach
developed in Ref. [15] and described here, one can overcome this difficulty.
 
Another interesting cosmological test for our model is the study of the
$M/L$ ratio for clusters of galaxies. The most recent average
value\cite{Bahcall}, $\Omega_{m} = 0.17 \pm 0.05$, has been extracted from the observation
of 21 clusters with $z \sim 1$. The fact that nearby cluster data seem to prefer smaller
values for $\Omega_{m}$ than the one obtained from WMAP data, can be interpreted 
as a signal in favour of a decaying dark matter model such as the GCG.


\section{Conclusions}

In this contribution, we have presented a setup where the GCG has been decomposed in two
interacting components.  The first one behaves as dark matter
since it is pressure-less. The second one has an equation of state,
$p_X = \omega_X \rho_X$. It has been shown that $\omega_X \le -1$.
Thus, once phantom-like behaviour is excluded the decomposition is
unique. Apparently the model does not look different from the
interacting quintessence models where one has two different interacting fluids; 
however, an interesting feature of our proposal is that it can be described through 
a single fluid equation. Hence, as far
the background cosmology is concerned, we have an unified GCG fluid
behaving as dark matter in the past and as a dark energy in the
present. Nevertheless, when studying structure formation in this model
one should consider it as an interacting mixture of two fluids to achieve a proper
description. In any unified model, one expects an entangled
mixture of interacting dark matter and dark energy. 
In the case of the GCG, we can uniquely identify the
components of this mixture and the interaction. Moreover, we find that
one does not need anything besides an evolving cosmological term to describe dark energy. 
This is consistent with recent studies that show that a
combination of WMAP data and observations of high redshift supernovae
can be described via a cosmological constant-like dark
energy\cite{Hari}. One can also consider the GCG as a decaying dark matter model
where the decay product is a cosmological constant. 

Obviously it remains to be seen how one can obtain such a decaying dark matter 
model from a fundamental theory. Given the fact that the GCG  equation of
state  arises from a generalized Born-Infeld action, it is possible that 
D-brane physics can shed some light into this issue (see eg. Ref. [24]).

Furthermore, we have demonstrated that in the context of our setup, the so-called dark
energy dominance is related with the time when matter fluctuations
become large ($\delta_{dm} > 1$), a possibility has actually been previously conjectured\cite{Varun}.
Moreover, we have shown that in
what concerns structure formation, the linear regime ($\delta_{dm}
\sim a$) is valid till fairly close to the present, meaning that at the
time structure formation begins, $z_c \simeq 10$, the influence of the dark energy
component was negligible and that clustering occurs very much like in
the CDM model.  We have shown that the growth factor as well as the
bias parameter have a noticeable dependence on the $\alpha$
parameter. We have implemented a model which exhibits a violation of
the Equivalence Principle, as dark energy and baryons are not directly
coupled. This may turn out to be an important observational signature of
our approach.

\section*{Acknowledgments}
It is a pleasure to thank Maria Bento, Anjan Sen, Somasri Sen and Pedro Silva for 
sharing the fun on the research of the GCG properties.


\end{document}